\documentclass[english,tightenlines,eqsecnum,amsmath,amssymb,aps,prd,superscriptaddress,nofootinbib,showpacs,floats]{article}
\usepackage{amsmath,amsfonts,amssymb,multicol}
\usepackage{graphicx,caption,subcaption}
\usepackage{epstopdf}
\usepackage[usenames]{xcolor}
\usepackage{epstopdf}
\definecolor{AHZ}{rgb}{0.0,1,0.0}
\usepackage[colorlinks,linkcolor=AHZ,citecolor=red,bookmarks]{hyperref}
\usepackage{rotating}
\usepackage[mathscr]{eucal}
\usepackage{graphicx}
\usepackage[T1]{fontenc}
\usepackage{babel}
\usepackage{authblk}
\usepackage{epsfig}
\usepackage{color}
\usepackage{amssymb}
\usepackage{fancyhdr}

\def\nn{\nonumber\\}
\newcommand{\f}[2]{\frac{#1}{#2}}
\def\be{\begin{equation}}
\def\ee{\end{equation}}
\def\bea{\begin{eqnarray}}
\def\eea{\end{eqnarray}}

\def\bwt{\begin{widetext}}
	\def\ewt{\end{widetext}}

\usepackage{amsmath}

\usepackage{amssymb}

\begin{document}
	
	\title{Charged wormhole solutions in Einstein-Cartan gravity}
	\author[1]{Mohammad Reza Mehdizadeh\thanks{mehdizadeh.mr@uk.ac.ir}}
	\author[2]{Amir Hadi Ziaie\thanks{ah.ziaie@riaam.ac.ir}}
	\affil[1]{{\rm Department~of~ Physics,~ Shahid~ Bahonar~ University, P.O.~ Box~ 76175, Kerman, Iran}}
	\affil[2]{{\rm Research~Institute~for~Astronomy~and~Astrophysics~of~ Maragha~(RIAAM), P.O.~Box~55134-441,~Maragha,~Iran}}
	\renewcommand\Authands{ and }
	\maketitle
\begin{abstract}
Static solutions representing wormhole configurations in Einstein-Cartan theory ({{\sf ECT}}) in the presence of electric charge are obtained. The solutions are described by a constant redshift function with matter content consisting of a Weyssenhoff fluid along with an anisotropic matter and energy momentum tensor ({\sf EMT}) of the electric field which together generalize the anisotropic energy momentum tensor in Einstein-Maxwell theory in order to include the effects of the intrinsic angular momentum (spin) of the particles. Assuming the equation of state ({{\sf EoS}}) $p_r={\sf w}_1\rho$ and $p_t={\sf w}_2\rho$, we derive exact wormhole solutions satisfying weak and null energy conditions. Depending on the value of the spin square density at the wormhole throat these solutions can be asymptotically flat, de-Sitter or anti de-Sitter. Observational aspects of the wormhole solutions are also studied, using gravitational lensing effects. It is found that the throat can act as a photon sphere near which the light deflection angle has arbitrarily large values. Moreover, for a particular class of solutions, when ${\sf w}_1\rightarrow-{\sf w}_2$ the lensing features of the present model mimic those of the Ellis wormhole in the weak field limit.
\end{abstract}
\maketitle
\section{Introduction}
Wormholes are topological handles that connect two spacetimes of the same Universe (as a bridge or tunnel) or of different Universes together by a minimal surface called the throat of the wormhole. This surface respects the flare-out condition \cite{misner-wheeler} through which a traveler can freely pass in both directions. 
The concept of wormhole was born in the seminal works of Misner and Wheeler \cite{misner-wheeler} and Wheeler \cite{Wheelerworm} in order to present a mechanism for having electric or magnetic \lq\lq{}charge without charge\rq\rq{} by letting the lines of force thread from one spatial asymptotic to another. The most amazing feature of a wormhole configuration is its two-way traversability which happens when the throat remains open. Therefore, it is important to recognize the possibility of traveling across the wormhole as a shortcut in spacetime. Unfortunately, a Schwarzschild wormhole does not possess this property and it is nontraversable, even by a photon~\cite{FulWheel}. This issue was investigated in a pioneering work by Morris and Thorne~\cite{mt} and subsequently Morris, Thorne and Yurtsever~\cite{mt1} where they introduced a static spherically symmetric metric and discussed the required conditions for physically meaningful Lorentzian traversable wormholes. However, the possibility for a wormhole to be traversable leads inevitably to violation of null energy condition ({\sf NEC}). In other words, the matter yielding this geometry is known as exotic; i.e. its energy density becomes negative which results in violation of {\sf NEC}~\cite{khu}. The quest for finding the promising candidates of exotic matter is not an easy task, and the footprints of exotic matter have been recognized only in a small area, such as the quantum Casimir effect and semiclassical Hawking radiation. However, all classical matter fields respect the standard energy conditions.
\par
By increasing the accuracy of measurements in observational cosmology and discovery of cosmic acceleration of the Universe, different cosmological models of dark energy with exotic equations of state have been put forward~\cite{exoicmattercos}. Among these models, possible traversable wormhole geometries can be built by matter fields with exotic {\sf EMT}~\cite{MarcoChianese2017}, phantom or quintom-type energy~\cite{phantworm} and interacting dark sectors~\cite{intdarksec}. One of the most important challenges in construction wormhole geometries is the fulfillment of standard energy conditions and instead of considering nonstandard fluids, many attempts have been done toward modifying general relativity ({\sf GR}) in order to overcome the issue of energy conditions within wormhole settings. In this regard, the study of wormhole solutions has recently attracted many people in modified theories of gravity e.g., the presence of higher order terms in curvature would allow for building thin-shell wormholes supported by ordinary matter \cite{thi}. A good deal of work along this line has been done in order to build and study wormhole solutions without resorting to exotic matter, among which we can quote: wormhole solutions in higher dimensional Lovelock theories~\cite{LOVEWORM}, Rastall gravity~\cite{rastallworm}, scalar-tensor theory \cite{bd}, $f({\sf R})$ gravity \cite{fr}, Einstein-Gauss-Bonnet theory~\cite{gmfl} and other theories.
\par
The Einstein-Cartan theory of spacetime is motivated by the desire to give a simple description of the influence of intrinsic angular momentum of microscopic matter (spin of fermionic particles) within gravitational phenomena. This objective can be achieved by taking the sapcetime as a four-dimensional differential manifold endowed with a metric tensor and a linear connection which in general is asymmetric. The spacetime torsion tensor is defined as the antisymmetric part of the connection and is physically generated through the presence of  spin of fermionic matter fields. The field equations of {\sf ECT} relate certain combinations of the curvature and torsion tensors to the energy-momentum and spin density tensors, respectively. Thus, in {\sf ECT}, both mass and spin, which are intrinsic and fundamental properties of matter fields would affect the spacetime structure. The essential idea behind {\sf ECT} was advanced by Cartan in early 1923 and further developed by Sciama and Kibble, 
see \cite{ECT3} for a beautiful and comprehensive review. Since the advent of {\sf ECT}, various cosmological as well as astrophysical models have been proposed with the aim of explaining the observed accelerated expansion of the Universe and also curing the problem of spacetime singularities which are unavoidable in {\sf GR}~\cite{hawellis}. Cosmological models in {\sf ECT} have been investigated in the last decades and it is shown that the spacetime torsion may provide a framework by virtue of which the initial singularity of the Universe is replaced by a nonsingular bounce \cite{spin-bounce}. Work along this line has been also extended to study the effects of spin in the early Universe~\cite{Gas}, the gravitational collapse scenario, and black hole physics~\cite{bhphystor}; see also~\cite{putzfield} for a careful collection of articles on different aspects of {\sf ECT}. Recently, static solutions in {\sf ECT} representing traversable wormholes with flat or anti-de Sitter (AdS) asymptotic behavior have been studied in~\cite{Broni-twoscalarfield}, where the matter sources are considered as two noninteracting scalar fields (one is minimally and the other is nonminimally coupled to gravity) with nonzero potentials. The obtained wormhole solutions also satisfy the {\sf NEC} and weak energy condition ({\sf WEC}) with arbitrary throat radius without resorting to exotic matter sources. Moreover, exact wormhole spacetimes with sources in the form of a nonminimally coupled nonphantom scalar field and an electromagnetic field have been found in \cite{Broniprd2016}. Further work in this area has been performed in~\cite{mehdihadi} where exact asymptotically flat and AdS spacetimes were obtained which admit traversable wormholes and respect energy conditions. The inclusion of charge within wormhole configurations was considered as a possibility to meet some stability conditions for wormhole geometry~\cite{KIMSWORM}, though the authors were not concerned with the fulfillment of energy conditions.  In the present paper we are interested in finding charged wormhole solutions in {\sf ECT} with supporting matter fields including a charged spinning fluid together with
an anisotropic ordinary matter distribution. We then proceed, in section~\ref{EC}, with introducing the field equations of {\sf ECT} with additional part in material components, i.e., the {\sf EMT} of the Maxwell field. In section~\ref{WHS1}, assuming that the radial and tangential pressures linearly depend on energy density via two {\sf EoS} parameters, we find exact wormhole solutions with zero tidal force satisfying {\sf WEC} and {\sf NEC}. Section~\ref{OBSFEATURE} deals with observational features of the wormhole solutions and our conclusions are drawn in section~\ref{concluding}.
\section{Field equations in Einstein-Cartan theory with a charged source}\label{EC}
The field equations in {\sf ECT} are given by \cite{ECT3,Venzo,mehzia}
\bea\label{ecfieldeq}
{\sf G}_{\mu\beta}\left(\{\}\right)-\Lambda{\sf  g}_{\mu\beta}=\kappa^2\left({{\sf T}}_{\mu\beta}+\theta_{\mu\beta}\right),~~~{\sf Q}^{\alpha}_{~\mu\beta}=-\f{\kappa^2}{2}\left[\Sigma_{\mu\beta}^{\,\,\,\,\,\,\,\alpha}+\f{1}{2}\delta^{\alpha}_{\,\,\mu}\Sigma_{\beta\rho}^{\,\,\,\,\,\,\,\rho}-\f{1}{2}\delta^{\alpha}_{\,\,\beta}\Sigma_{\mu\rho}^{\,\,\,\,\,\,\,\rho}\right],
\eea
where $\kappa^2=8\pi G/c^4$ and $\Lambda$ are the gravitational coupling and cosmological constants, ${\sf Q}^{\alpha}_{~\mu\beta}$ is the spacetime torsion tensor and $\Sigma^{\mu\alpha\beta}$ is defined as the spin tensor of matter \cite{ECT3}. The tensor $\theta_{\mu\beta}$ represents a correction due to the spin contributions to the dynamical {\sf EMT}, i.e., ${{\sf T}}_{\mu\beta}$~\cite{spfieldspop}, and can be obtained in terms of torsion tensor (or equivalently spin tensor) as
\bea\label{correctionterms}
\theta_{\mu\nu}&=&\f{1}{\kappa^2}\Bigg[4{\sf Q}^{\eta}_{\,\,\mu\eta}{\sf Q}^{\beta}_{\,\,\nu\beta}-\left({\sf Q}^{\rho}_{\,\,\,\mu\epsilon}+2{\sf Q}_{(\mu\epsilon)}^{\,\,\,\,\,\,\,\rho}\right)\left({\sf Q}^{\epsilon}_{\,\,\nu\rho}+2{\sf Q}_{(\nu\rho)}^{\,\,\,\,\,\,\,\,\epsilon}\right)+\f{1}{2}{\sf g}_{\mu\nu}\left({\sf Q}^{\rho\sigma\epsilon}+2{\sf Q}^{(\sigma\epsilon)\rho}\right)\left({\sf Q}_{\epsilon\sigma\rho}+2{\sf Q}_{(\sigma\rho)\epsilon}\right)\nonumber\\&-&2{\sf g}_{\mu\nu}{\sf Q}^{\rho\epsilon}_{\,\,\,\,\rho}{\sf Q}^{\sigma}_{\,\,\,\epsilon\sigma}
\Bigg]\nn
&=&\f{1}{2}\kappa^2\bigg[\Sigma_{\mu\alpha}^{~~\,\alpha}\Sigma_{\nu\gamma}^{~~~\!\!\gamma}-\Sigma_{\mu}^{~\alpha\gamma}\Sigma_{\nu\gamma\alpha}-\Sigma_{\mu}^{~\alpha\gamma}\Sigma_{\nu\alpha\gamma}\nn&+&\f{1}{2}\Sigma^{\alpha\gamma}_{~~\,\mu}\Sigma_{\alpha\gamma\nu}+\f{1}{4}{\sf g}_{\mu\nu}\left(2\Sigma_{\alpha\gamma\epsilon}\Sigma^{\alpha\epsilon\gamma}
-2\Sigma_{\alpha~\gamma}^{~\gamma}\Sigma^{\alpha\epsilon}_{~~\,\epsilon}
+\Sigma^{\alpha\gamma\epsilon}\Sigma_{\alpha\gamma\epsilon}\right)\bigg],\eea
where use has been made of the second part of (\ref{ecfieldeq}) and $()$ denotes symmetrization. It should be mentioned that the equation governing the torsion tensor is purely algebraic. This means that the torsion does not propagate outside the matter distribution as a torsion wave or through any interaction of nonvanishing range~\cite{ECT3} and, therefore, is only nonzero inside the matter source. Next, we proceed to find a suitable description for {\sf EMT} given in (\ref{correctionterms}) in terms of a spin fluid. Such a fluid can be described by the so-called Weyssenhoff fluid considered as a continuous macroscopic medium whose microscopic elements are composed of fermionic particles with intrinsic angular momentum. This model which generalizes the {\sf EMT} of ordinary matter in {\sf GR} to include nonvanishing spin was first studied by Weyssenhoff and Raabe \cite{W1947} and extended by Obukhov and Korotky in order to construct cosmological models based on the {\sf ECT}~\cite{KCQG1987}. In order to consider wormhole solutions in the framework of {\sf ECT}, we use a classical description of spin as postulated by Weyssenhoff given by~\cite{W1947},\cite{KCQG1987},
\be\label{FC}
\Sigma_{\mu\nu}^{~~\alpha}={\sf s}_{\mu\nu}{\rm u}^{\alpha},~~~~~~~~{\sf s}_{\mu\nu}{\rm u}^{\mu}=0,
\ee 
where ${\rm u}^{\alpha}$ is the four-velocity of the fluid element and ${\sf s}_{\mu\nu}=-{\sf s}_{\nu\mu}$ is a second-rank antisymmetric tensor defined as the spin density tensor. The spatial components of spin density tensor include the 3-vector\footnote{We use the convention $(t, r, \theta, \phi) = (0, 1, 2, 3)$ for labeling the coordinates.} $({\sf s}^{23},{\sf s}^{13},{\sf s}^{12})$ which coincides in the rest frame with the spatial spin density of matter element. The rest of the spacetime components $({\sf s}^{01}, {\sf s}^{02},{\sf s}^{03})$ are assumed to be zero in the rest frame of the fluid element, which can be covariantly formulated as a constraint given in the second part of (\ref{FC}). As we are concerned with a charged spinning fluid, we consider a dynamical {\sf EMT} which includes three parts i.e., the usual perfect fluid part, ${\sf T}^{\sf Pf}_{\mu\beta}$, an intrinsic spin part ${\sf T}^{\sf s}_{\mu\beta}$ and contributions due to electromagnetic tensors, ${\sf T}^{{\sf EM}}_{\mu\beta}$. We, therefore, have \cite{Gas,W1947,KCQG1987}
\bea\label{sfss}
{{\sf T}}_{\alpha\beta}&=&{\sf T}^{\sf Pf}_{\alpha\beta}+{\sf T}^{\sf s}_{\alpha\beta}+{\sf T}^{{\sf EM}}_{\alpha\beta}=\left\{(\rho+p_{t}){\rm u}_\alpha {\rm u}_\beta+p_{t}{\sf g}_{\alpha\beta}+(p_r-p_t){\rm v}_\alpha {\rm v}_\beta\right\}\nonumber\\&+&{\rm u}_{(\alpha}{\sf s}_{\beta)}^{\,\,\,\mu}{\rm u}^{\nu}{{\sf K}}^{\rho}_{\,\,\mu\nu}{\rm u}_{\rho}+{\rm u}^{\rho}{{\sf K}}^{\mu}_{\,\,\rho\sigma}{\rm u}^{\sigma}{\rm u}_{(\alpha}{\sf s}_{\beta)\mu}-\f{1}{2}{\rm u}_{(\alpha}{{\sf Q}}_{\beta)\mu\nu}{\sf s}^{\mu\nu}
+\f{1}{2}{{\sf Q}}_{\nu\mu(\alpha}{\sf s}^{\mu}_{\,\,\beta)}{\rm u}^\nu,\nn
&+&\frac{1}{4\pi}\left[{\sf F}_{\alpha}^{\,\,\,\nu}{\sf F}_{\beta\nu}-\f{{\sf g}_{\alpha\beta}}{4}{\sf F}_{\mu\nu}{\sf F}^{\mu\nu}\right],
\eea
where ${\rm v}_{\mu}$ is a unit spacelike vector field in radial direction and quantities $\rho$, $p_r$ and $p_t$ are the usual energy density, radial and tangential pressures of the fluid, respectively. The quantity ${\sf K}^{\mu}_{\,\,\,\nu\alpha}$ is the contorsion tensor defined as
\be\label{contortion}
{\sf K}^{\mu}_{~\alpha\beta}={\sf Q}^{\mu}_{~\alpha\beta}+{\sf Q}_{\alpha\beta}^{~~\,\mu}+
{\sf Q}_{\beta\alpha}^{~~\,\mu}.
\ee
The electromagnetic field tensor
\be\label{emfield}{\sf F}_{\mu\nu}=\partial_{\mu}{\sf A}_\nu-\partial_{\nu}{\sf A}_\mu,
\ee
with ${\sf A}_{\mu}$ being the electromagnetic potential is an antisymmetric tensor field satisfying the Maxwell field equations\footnote{As has been pointed out by \cite{emcouptor,Hehlgrg}, the electromagnetic field does not couple with the spacetime torsion thus, the Maxwell field equations (\ref{maxfieldeqs}) are written in their usual way. If the electromagnetic field couples to torsion the gauge invariance is broken.}
\be\label{maxfieldeqs}
\partial_{\mu}{\sf F}_{\alpha\nu}+\partial_{\nu}{\sf F}_{\mu\alpha}+\partial_{\alpha}{\sf F}_{\nu\mu}=0,~~~~~\partial_\mu\left[(-{\sf g})^{\f{1}{2}}{\sf F}^{\mu\nu}\right]=(-{\sf g})^{\f{1}{2}}{\sf J}^\nu,
\ee
where ${\sf g}$ is the metric determinant and ${\sf J}^{\nu}$ is the current four-vector defined via the proper charge density as
\be\label{JV}
{\sf J}^\nu=\sigma(r){\rm u}^{\nu}.
\ee

\section{Wormhole Solutions}\label{WHS1}
\begin{figure}
\begin{center}
\includegraphics[scale=0.35]{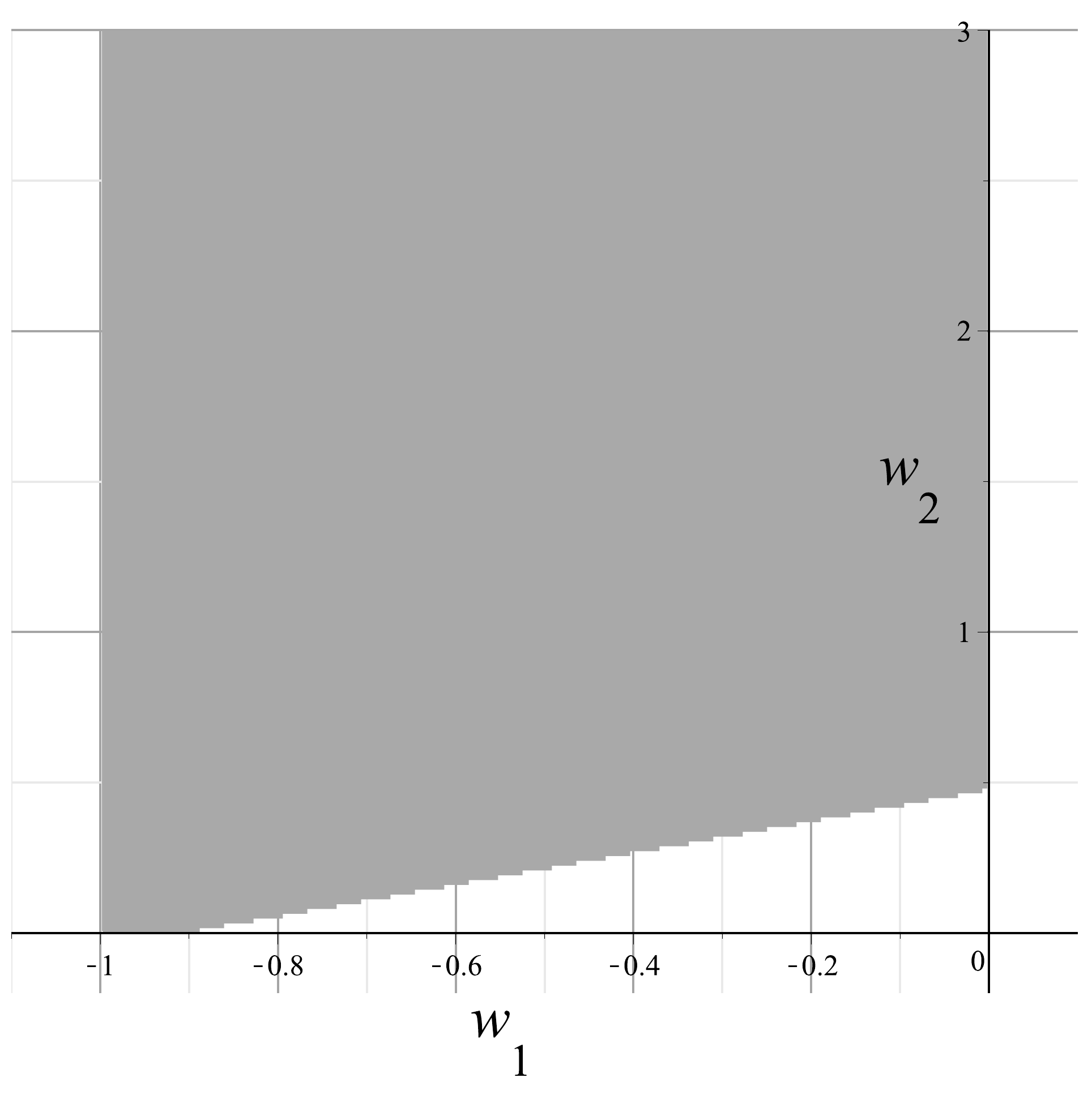}
\caption{The allowed values of {\sf EoS} parameters for ${\sf Q}_0=1.1$, $\Lambda=0$ and $r_0=1$.}\label{fig1}
\end{center}
\end{figure}

\begin{figure}
\begin{center}
\includegraphics[scale=0.3]{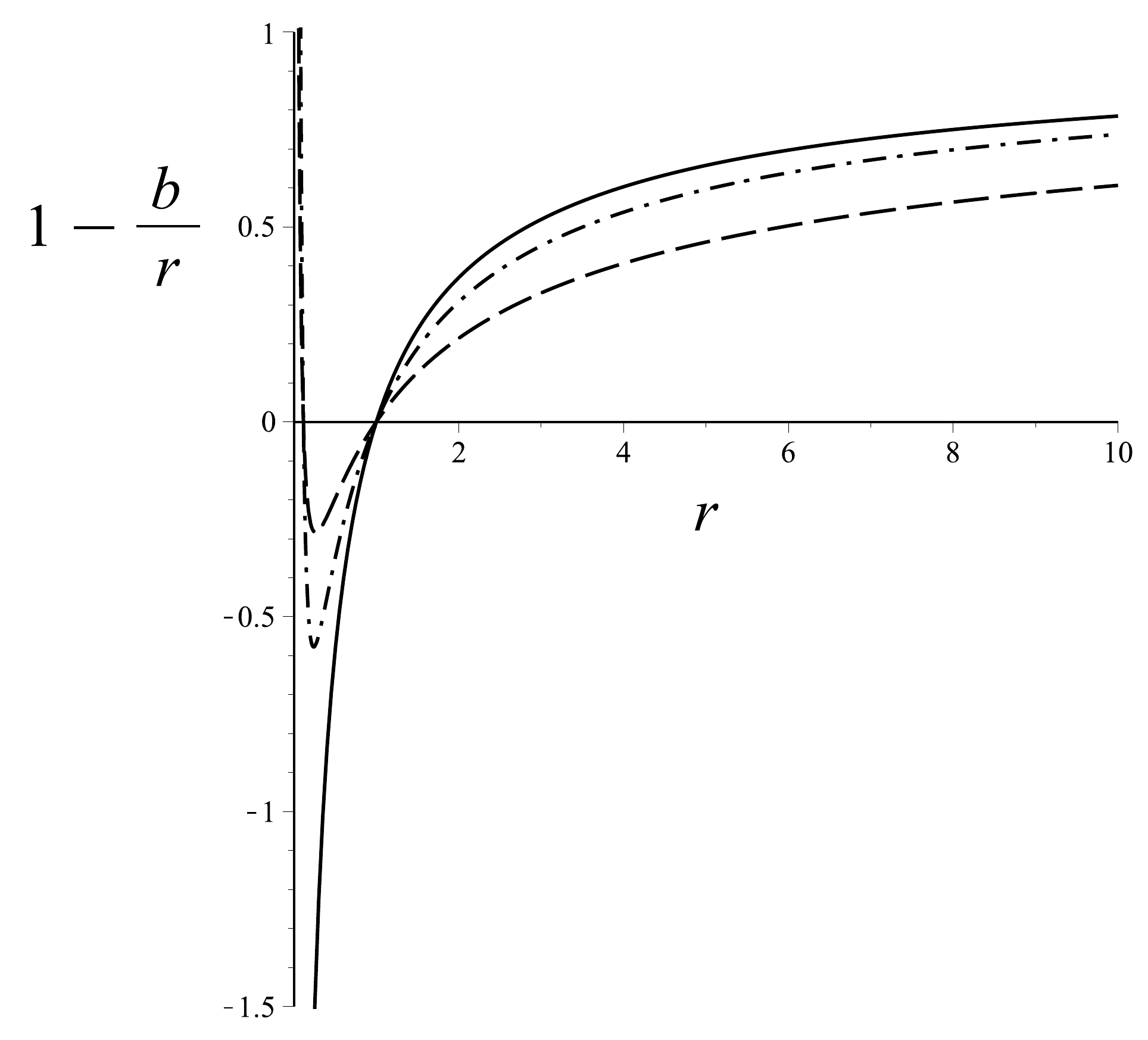}
\includegraphics[scale=0.28]{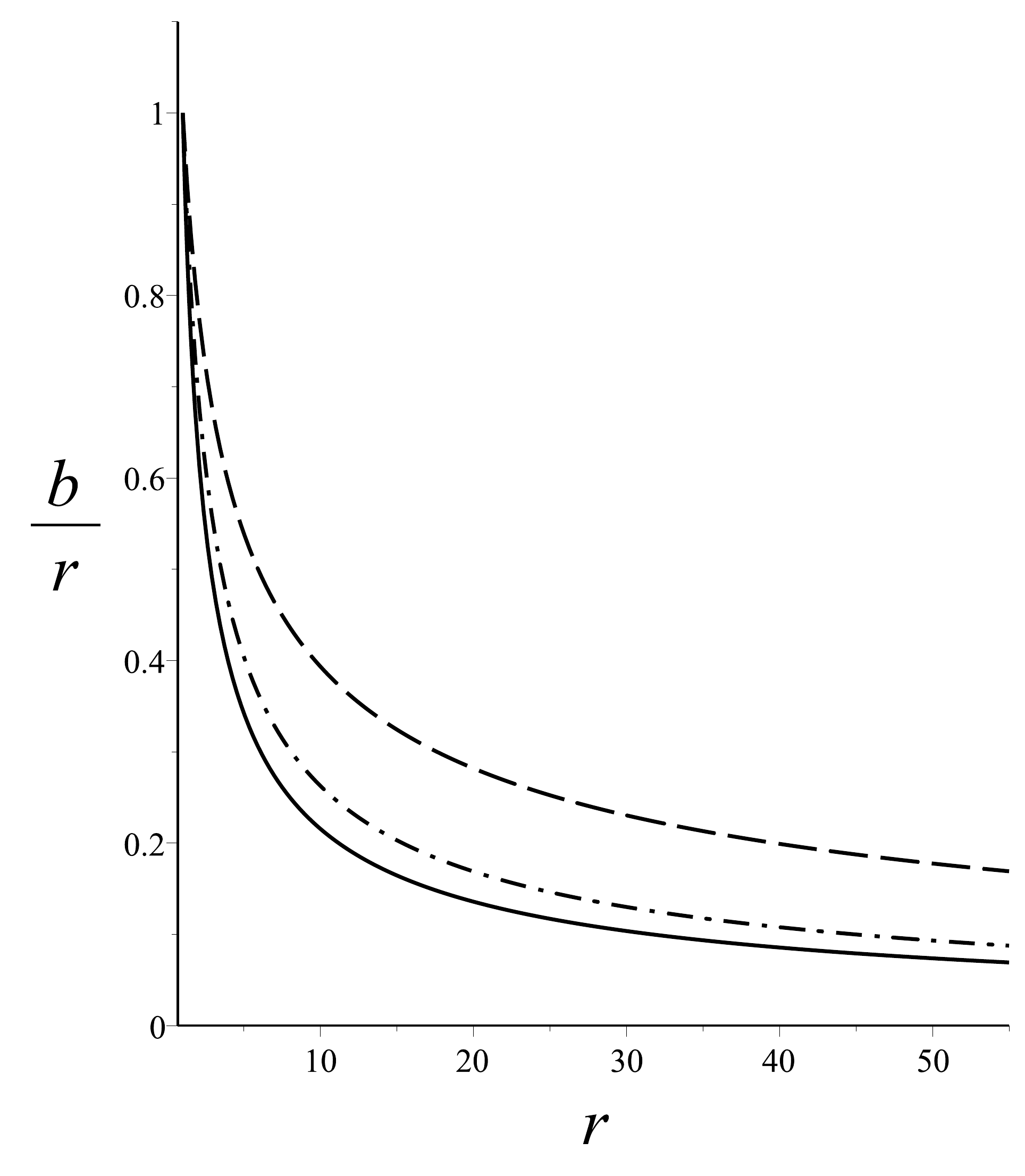}
\caption{Plot of ${\sf g}_{{\rm rr}}^{-1}(r)$ (left panel) and the ratio $b(r)/r$ (right panel) for ${\sf Q}_0=1.1$, $\Lambda=0$ and $r_0=1$, ${\sf w}_1=-0.33$, ${\sf w}_2=1$ (solid curve), ${\sf w}_1=-1$, ${\sf w}_2=0.33$ (dashed curve) and ${\sf w}_1=-1$, ${\sf w}_2=0.63$ (dot-dashed curve).}\label{fig2}
\end{center}
\end{figure}

\begin{figure}
\begin{center}
\includegraphics[scale=0.3]{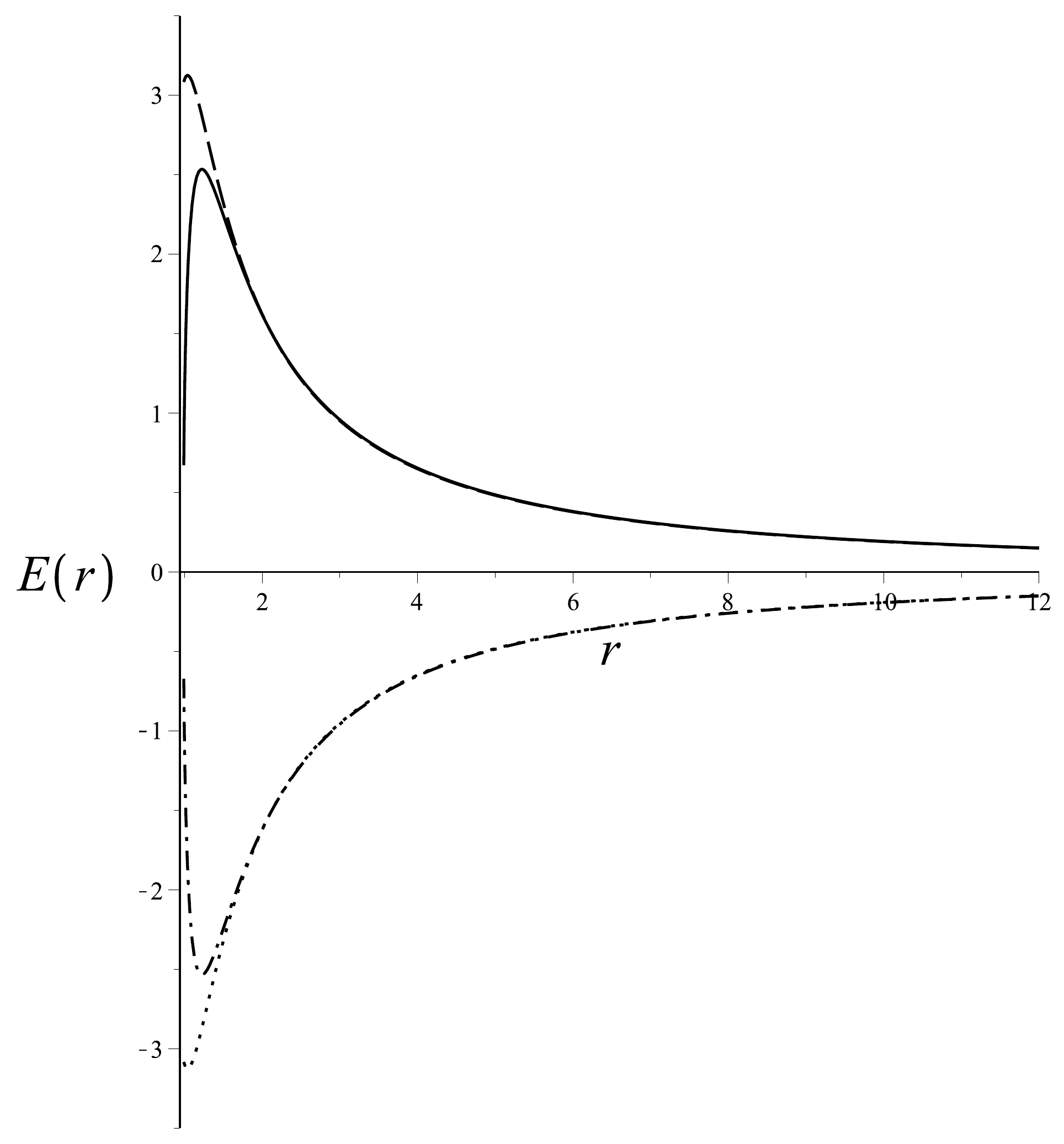}
\caption{The behavior of electric field for ${\sf w}_1=-0.3$, ${\sf w}_2=0.8$,  ${\sf Q}_0=1.1$ (solid curve) and ${\sf Q}_0=3.1$ (dashed curve). Dotted and dot-dashed curves are plotted for the same values but negative sign of (\ref{EXACTSOL}). We have set $\Lambda=0$ and $r_0=1$.}\label{fig4}
\end{center}
\end{figure}

\begin{figure}
\begin{center}
\includegraphics[scale=0.27]{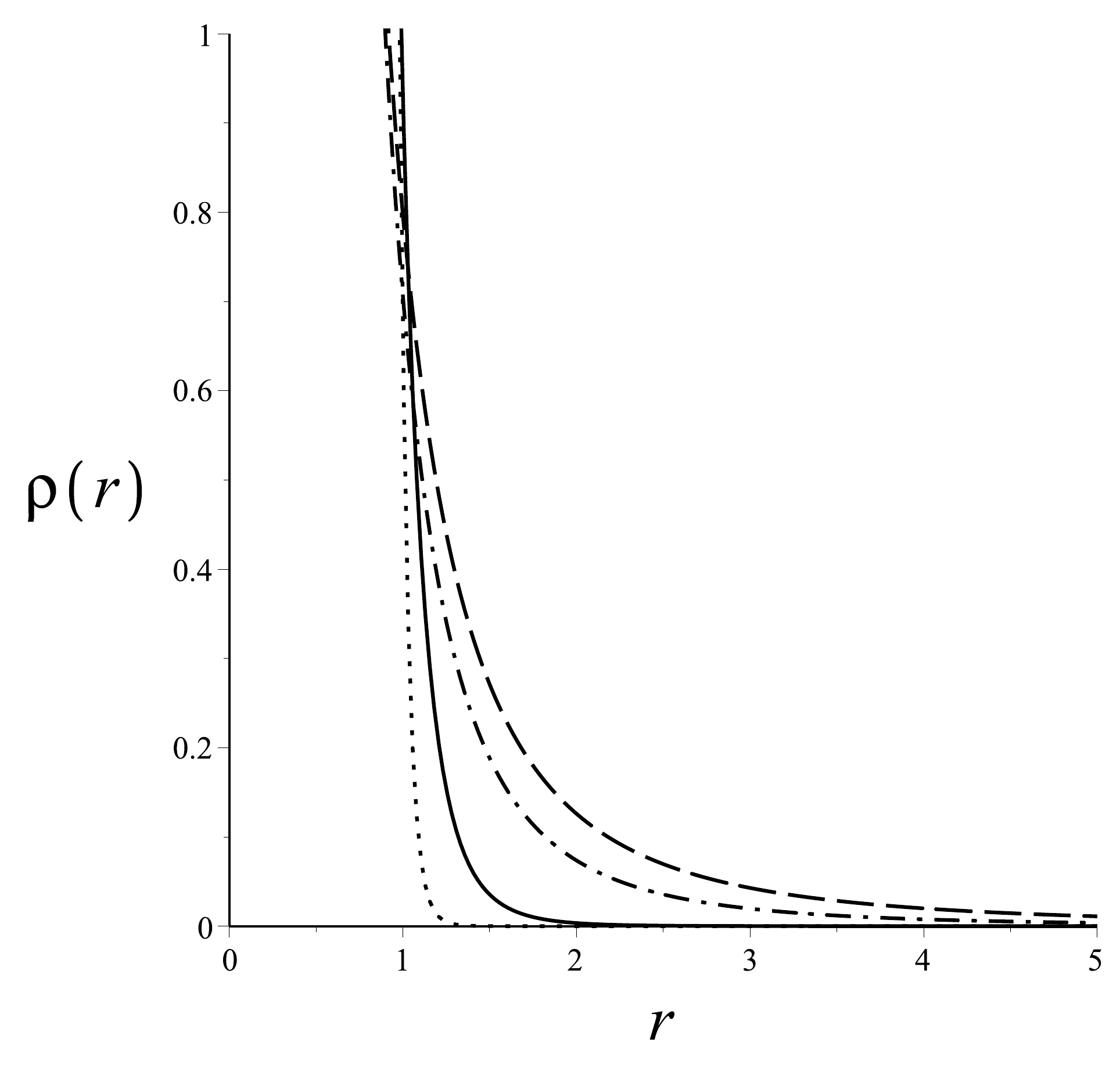}
\includegraphics[scale=0.27]{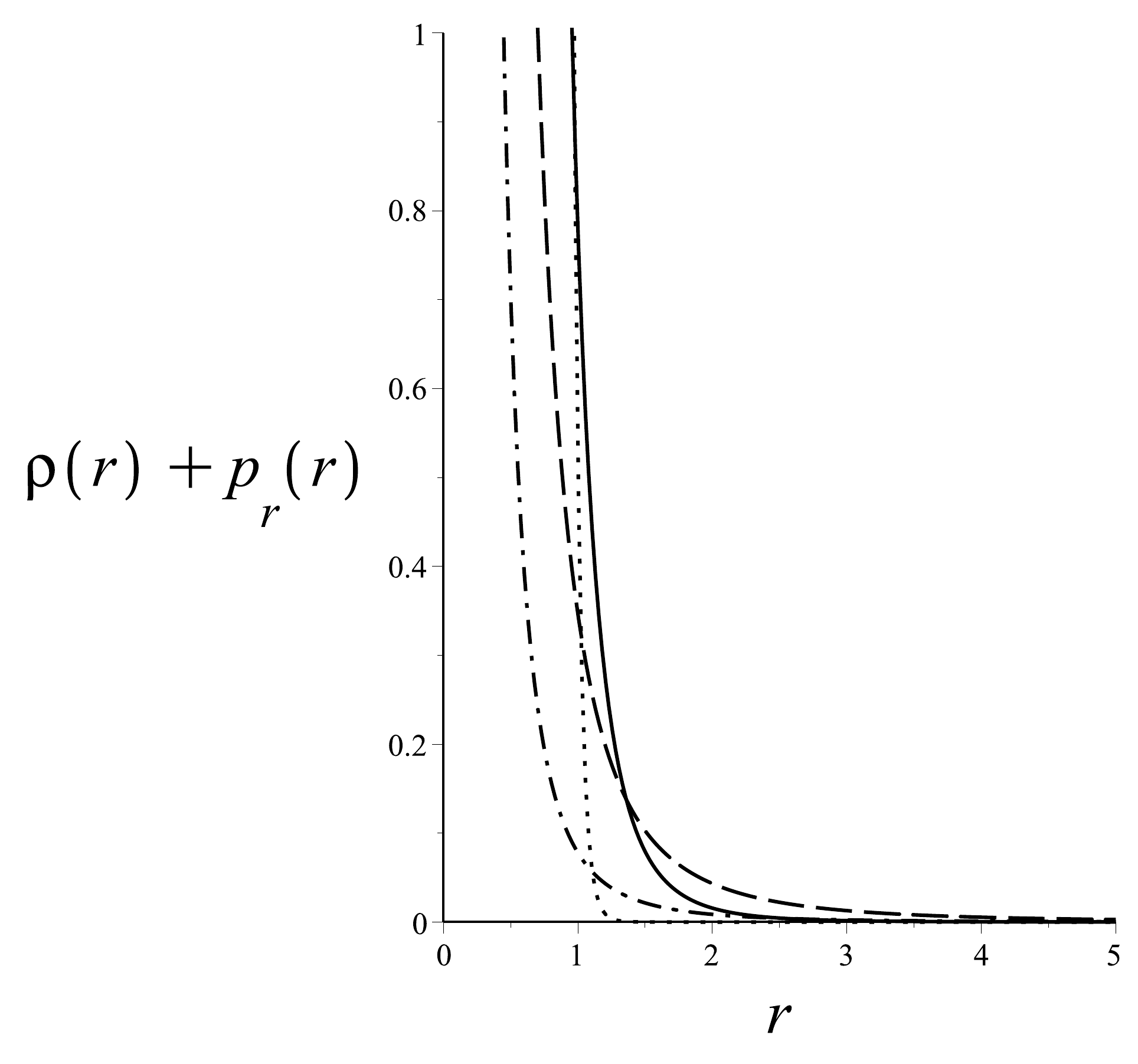}
\includegraphics[scale=0.27]{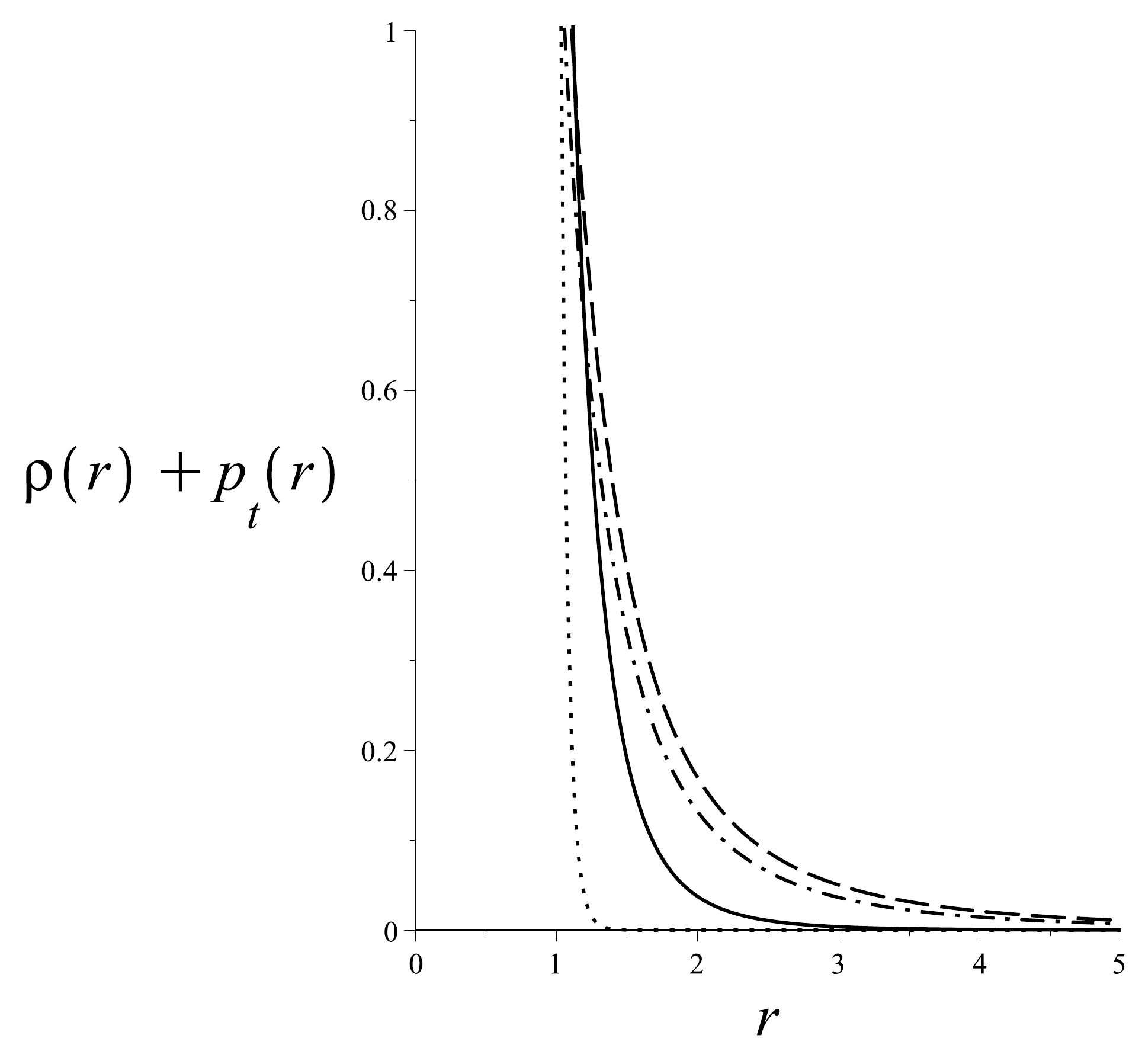}
\includegraphics[scale=0.27]{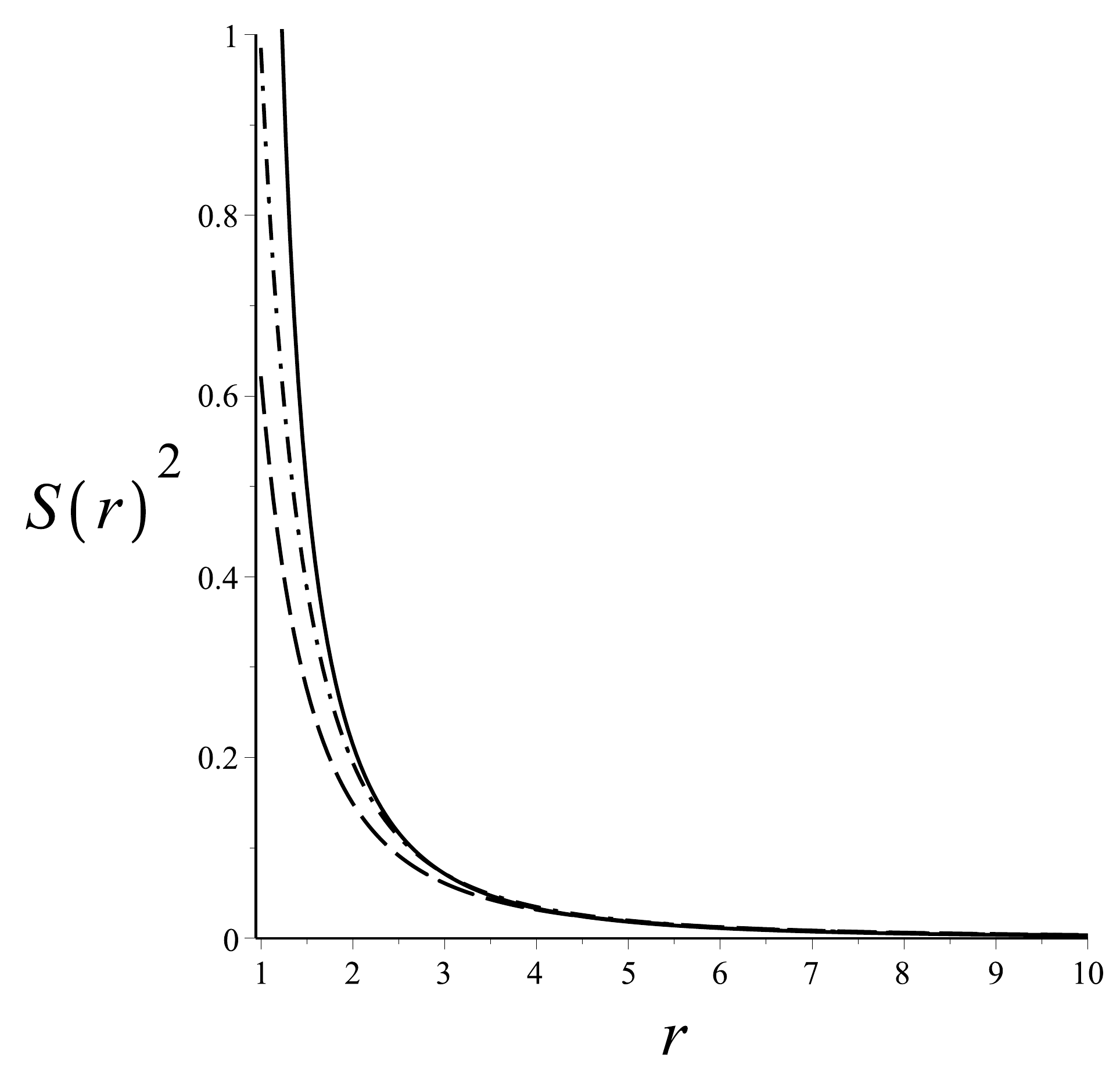}
\caption{The behavior of energy density (upper left), $\rho(r)+p_r(r)$ (upper right) and $\rho(r)+p_t(r)$ (lower left) for ${\sf Q}_0=1.1$, $\Lambda=0$, $r_0=1$, ${\sf w}_1=-0.33$, ${\sf w}_2=1$ (solid curve), ${\sf w}_1=-1$, ${\sf w}_2=0.33$ (dashed curve), ${\sf w}_1=-1$, ${\sf w}_2=0.63$ (dot-dashed curve), ${\sf w}_1=-0.2$, ${\sf w}_2=2$ (dotted curve). (Lower right) The behavior of square of spin density for ${\sf Q}_0=1.1$, $\Lambda=0$, $r_0=1$, ${\sf w}_1=-0.33$, ${\sf w}_2=1$ (solid curve), ${\sf w}_1=-1$, ${\sf w}_2=0.33$ (dashed curve), ${\sf w}_1=-1$, ${\sf w}_2=0.63$ (dot-dashed curve).}\label{fig6}
\end{center}
\end{figure}

We consider the general static and spherically symmetric line element representing a wormhole given by (we set the units so that $\kappa=1$)
\begin{eqnarray}\label{evw}
ds^2=-{\rm e}^{2\Phi(r)}dt^2+\left(1-\f{b(r)}{r}\right)^{-1}dr^2+r^2d\Omega^2,
\end{eqnarray}
where $d\Omega^2=d\theta^2+\sin^2\theta d\phi^2$ is the standard line element on a unit two-sphere, $\Phi(r)$ is the redshift function and $b(r)$ is the wormhole shape function. The radial coordinate has a range so that it increases from a minimum value at $r_0$ (wormhole\rq{}s throat) to spatial infinity. Conditions on $\Phi (r)$ and $b(r)$ under which wormholes are traversable
were discussed completely for the first time in \cite{mt}. The shape function must satisfy the flare-out condition at the throat; i.e., we must have $b^{\prime}(r_0)<1$ and $b(r)<r$ for $r>r_0$ in the whole spacetime. Our aim in the present work is to determine $b(r)$ and $\Phi(r)$ in order to construct physically reasonable wormhole geometries. Following \cite{prasana75,emcouptor} we suppose that the spins of the individual charged particles are all aligned in the radial direction. Therefore from (\ref{FC}) we obtain ${\sf s}_{23}=-{\sf s}_{32}={\sf S}$ as the only independent nonzero component of the spin density tensor. We then find the intrinsic angular momentum tensor of matter as
\be\label{spinangt}
\Sigma^{\,\,\,\ 0}_{23}=-\Sigma^{\,\,\,\ 0}_{32}={\sf S}({\sf g}_{00})^{-\f{1}{2}}.
\ee
In the present model the four-vector potential is given as
\be\label{VPA}
{\sf A}_\mu=\left[\Psi(r),0,0,0\right],
\ee
from which the electromagnetic field tensor is obtained as
\be\label{EMFTENA}
{\sf F}_{01}=-{\sf F}_{10}=\Psi^{\prime}.
\ee
The field equations (\ref{ecfieldeq}) and (\ref{maxfieldeqs}) then read
\bea
\rho(r)&=&\f{b^{\prime}}{r}+\f{{\sf S}^2}{4}-\f{{\sf E}^2}{8\pi}-\Lambda,\label{rhofeq}\\
p_r(r)&=&\f{2\Phi^\prime}{r}\left[1-\f{b}{r}\right]-\f{b}{r^3}+\f{{\sf E}^2}{8\pi}+\f{{\sf S}^2}{4}+\Lambda,\label{prfeq}\\
p_t(r)&=&\left[1-\f{b}{r}\right]\left(\Phi^{\prime\prime}+\Phi^{\prime2}\right)-\f{\Phi^{\prime}}{2r^2}\left[rb^\prime+b-2r\right]+\f{1}{2r^2}\left[\f{b}{r}-b^\prime\right]+\f{{\sf S}^2}{4}-\f{{\sf E}^2}{8\pi}+\Lambda,\label{ptfeq}\\
\sigma(r)&=&\f{1}{4\pi r^2}\left(1-\f{b}{r}\right)^{\f{1}{2}}(r{\sf E}^\prime+2{\sf E}),\label{sigfeq}
\eea
where the electric field strength (the ${\sf F}_{01}$ component of the electromagnetic field tensor ${\sf F}_{\mu\nu}$) is defined as~\cite{EFieldtiw}
\be\label{elecfield}
{\sf E}(r)=-\Psi^\prime{\sf exp}(-\Phi(r))\left[1-\f{b}{r}\right]^{\f{1}{2}},
\ee
and use has been made of ${\rm u}^\mu=\left[{\sf exp}(-\Phi(r)),0,0,0\right]$ and ${\rm v}_\mu=\left[0,\sqrt{1-b(r)/r},0,0\right]$. Equation (\ref{sigfeq}) can also be expressed in the following form
\be\label{EFieldQ}
{\sf E}(r)=\f{4\pi}{r^2}\int_{r_0}^r\f{\xi^2\sigma(\xi)}{\left[1-\f{b(\xi)}{\xi}\right]^{\f{1}{2}}}d\xi=\f{{\sf Q}(r)}{r^2},
\ee
where ${\sf Q}(r)$ is the total charge of the sphere of radius $r$.
The conservation equation 
\be\label{conseq}
-\Phi^\prime[\rho(r)+p_r(r)]-p_r^\prime+\f{2}{r}[p_t(r)-p_r(r)]=0,
\ee
leaves us with the following relation 
\begin{eqnarray}\label{CE1}
\f{{\sf S}}{2}\left[{\sf S}^\prime+{\sf S}\Phi^\prime\right]+\f{1}{4\pi}\left[{\sf E}{\sf E}^\prime+\f{2}{r}{\sf E}^2\right]=0,
\end{eqnarray}
or, equivalently,
\begin{eqnarray}\label{CE2}
\f{{\sf S}}{2}\left[{\sf S}^\prime+{\sf S}\Phi^\prime\right]+\f{1}{4\pi r^4}{\sf Q}{\sf Q}^\prime=0.
\end{eqnarray}
Considering $p_r=p_r(\rho)$ and $p_t=p_t(\rho)$, equations (\ref{rhofeq})-(\ref{ptfeq}) along with conservation equation (\ref{CE2}) constitute a system of differential equations to be solved for the unknowns. Let us assume that the radial and tangential components of the fluid pressure depend linearly on energy density, i.e., $p_r(r)={\sf w}_1\rho(r)$ and $p_t(r)={\sf w}_2\rho(r)$. Therefore, the anisotropic fluid behaves differently in radial and tangential directions depending on {\sf EoS} parameters, ${\sf w}_1$ and ${\sf w}_2$. Such an {\sf EoS} has been widely used in the literature for the study of wormhole configurations; see, e.g.~\cite{eosw1w2}. We then get
\bea\label{SYSEQS}
&&\f{b}{r^3}+\f{{\sf w}_1}{r^2}b^\prime-\f{2}{r}\left[1-\f{b}{r}\right]\Phi^\prime-\f{({\sf w}_1+1)}{8\pi r^4}{\sf Q}^2+\f{1}{4}({\sf w}_1-1){\sf S}^2-\Lambda({\sf w}_1+1)=0,\\
&&-\left[1-\f{b}{r}\right]\left(\Phi^{\prime\prime}+(\Phi^\prime)^2\right)+\f{\Phi^\prime}{2r^2}(rb^\prime-2r+b)+\f{b^\prime}{2r^2}(2{\sf w}_2+1)-\f{b}{2r^3}\nn
&+&\f{{\sf Q}^2}{8\pi r^4}(1-{\sf w}_2)+\f{{\sf S}^2}{4}({\sf w}_2-1)-\f{\Lambda}{8\pi r^4}({\sf w}_2+1)=0,\\
&&\f{{\sf S}}{2}\left[{\sf S}^\prime+{\sf S}\Phi^\prime\right]+\f{1}{4\pi r^4}{\sf Q}{\sf Q}^\prime=0.
\eea
The above system of differential equations is closed once we specify one of its four unknowns. As we know, one of the features of a traversable wormhole is that the tidal gravitational forces as experienced by a passenger must be reasonably small. We then proceed with a constant redshift function, i.e., $\Phi(r)=\Phi_0$ implying wormholes with zero tidal forces. We find out that the system admits an exact solution given by
\bea
{\sf Q}(r)&=&\f{\pm1}{{\sf w}_1+3{\sf w}_2}\left[3{\sf w}_1({\sf w}_1+3{\sf w}_2)\left({\sf C}_2r^{\f{4}{3}}-{\sf C}_1r^{\f{2({\sf w}_1+{\sf w}_2)}{{\sf w}_1}}\right)\right]^{\f{1}{2}},\label{EXACTSOL}\\
{\sf S}^2(r)&=&\f{3{\sf C}_2{\sf w}_1}{4\pi({\sf w}_1+3{\sf w}_2)r^{\f{8}{3}}}-\f{3{\sf C}_1{\sf w}_1({\sf w}_1+{\sf w}_2)}{2\pi({\sf w}_1-{\sf w}_2)({\sf w}_1+3{\sf w}_2)}r^{\f{2}{{\sf w}_1}({\sf w}_2-{\sf w}_1)}-2\Lambda,\label{EXACTSOL1}\\
b(r)&=&\f{1}{2}\Lambda r^3+\f{9{\sf C}_2{\sf w}_1}{16\pi({\sf w}_1+3{\sf w}_2)}r^{\f{1}{3}}+\f{3{\sf C}_1{\sf w}_1^2({\sf w}_2-1)}{4\pi({\sf w}_1+3{\sf w}_2)({\sf w}_1+2{\sf w}_2+1)({\sf w}_1-{\sf w}_2)}r^{\f{{\sf w}_1+2{\sf w}_2}{{\sf w}_1}},\label{EXACTSOL2}
\eea
where ${\sf C}_1$ and ${\sf C}_2$ are  integration constants. The first one can be determined subject to the condition that the shape function has to satisfy at the wormhole throat which is $b(r_0)=r_0$. This gives
\be\label{CC1}
{\sf C}_1= \f{2({\sf w}_1+2{\sf w}_2+1)({\sf w}_2-{\sf w}_1)\left[\f{9}{8}{\sf C}_2{\sf w}_1r_0^{\f{4}{3}}+\pi r_0^2({\sf w}_1+3{\sf w}_2)(\Lambda r_0^2-2)\right]}{3{\sf w}_1^2({\sf w}_2-1)r_0^{\f{2({\sf w}_1+{\sf w}_2)}{{\sf w}_1}}}.
\ee
The constant ${\sf C}_2$ can be found so that the charge function is finite at the wormhole throat. We then get
\be\label{chargeatth}
{\sf C}_2=\f{8\pi r_0^2\left(2-\Lambda r_0^2\right){\sf w}_1^2-\left[\pi r_0^2({\sf w}_2+1)(\Lambda r_0^2-2)-\f{1}{2}{\sf Q}_0^2({\sf w}_2-1)\right]{\sf w}_1+2\pi r_0^2(\Lambda r_0^2-2)\left({\sf w}_2+\f{1}{2}\right){\sf w}_2}{3{\sf w}_1(3{\sf w}_1-2{\sf w}_2-1)r_0^{\f{4}{3}}}.
\ee
Substituting for ${\sf C}_1$ and ${\sf C}_2$ back into the solution we arrive at the following relations for the charge function, square of the spin density, and shape function 
\bea
{\sf Q}(r)&\!\!\!=\!\!\!&\Bigg[4{\sf A}\left(\f{r}{r_0}\right)^{\f{4}{3}}+{\sf B}({\sf w}_1+2{\sf w}_2+1)({\sf w}_1-{\sf w}_2)\left(\f{r}{r_0}\right)^{\f{2({\sf w}_1+{\sf w}_2)}{{\sf w}_1}}\Bigg]^{\f{1}{2}},\label{finalsol00}\\
{\sf S}^2(r)&\!\!\!=\!\!\!&\f{{\sf A}}{\pi r_0^{4}}\left(\f{r_0}{r}\right)^{\f{8}{3}}+\f{({\sf w}_1+2{\sf w}_2+1)({\sf w}_1+{\sf w}_2){\sf B}}{2\pi r_0^4}\left(\f{r}{r_0}\right)^{\f{2({\sf w}_2-{\sf w}_1)}{{\sf w}_1}}-2\Lambda,\label{finalsol11}\\
\!\!\!b(r)&\!\!\!=\!\!\!&\f{1}{2}\Lambda r^3+\f{3{\sf A}}{4\pi r_0}\left(\f{r}{r_0}\right)^{\f{1}{3}}-\f{{\sf w}_1({\sf w}_2-1){\sf B}}{4\pi r_0}\left(\f{r}{r_0}\right)^{\f{{\sf w}_1+2{\sf w}_2}{{\sf w}_1}},\label{finalsol22}
\eea
where
\bea\label{coeff0}
{\sf A}&=&\f{{\sf Q}_0^2{\sf w}_1({\sf w}_2-1)-2\pi r_0^2( \Lambda r_0^2-2)\left({\sf w}_1-{\sf w}_2\right)({\sf w}_1+2{\sf w}_2+1)}{({\sf w}_1+3{\sf w}_2)(3{\sf w}_1-2{\sf w}_2-1)},\\
{\sf B}&=&\f{8\pi r_0^2(\Lambda r_0^2-2)+3{\sf Q}_0^2}{(3{\sf w}_1-2{\sf w}_2-1)({\sf w}_1+3{\sf w}_2)}.
\eea
From (\ref{finalsol22}) the flare-out condition ($b^\prime(r_0)<1$) gives
\be\label{ineqflare}
\f{\pi r_0^2\left[8\Lambda r_0^2({\sf w}_1-{\sf w}_2)+2{\sf w}_1+4{\sf w}_2-6\right]-{\sf Q}_0^2({\sf w}_2-1)}{\pi r_0^2(6{\sf w}_1-4{\sf w}_2-2)}<1.
\ee
We also take the value of the square of the spin density at the throat to be ${\sf S}^2(r_0)={\sf S}_0^2$, whereby we have
\be\label{lambdas0}
\Lambda=\f{-\pi r_0^2\left[{\sf S}_0^2r_0^2(3{\sf w}_1-2{\sf w}_2-1)+4{\sf w}_1+8{\sf w}_2+4\right]+{\sf Q}_0^2(3{\sf w}_1+2{\sf w}_2+1)}{4\pi r_0^2({\sf w}_1-2{\sf w}_2-1)}.
\ee
The solution (\ref{finalsol22}) is asymptotically flat for $\Lambda=0$, and ${\sf w}_1$ and ${\sf w}_2$ are of the opposite sign. For this case the square of spin density at the throat is found as
\be\label{s02}
{\sf S}_0^2=\f{(3{\sf w}_1+2{\sf w}_2+1){\sf Q}_0^2-8\pi r_0^2({\sf w}_1+2{\sf w}_2+1)}{2\pi r_0^4(3{\sf w}_1-2{\sf w}_2-1)}.
\ee
We note that equation (\ref{s02}) can also be solved for $r_0$ as a function of the model parameters. Hence, for specified {\sf EoS} parameters, the throat radius can admit only certain values determined by charge and spin square density at the throat. Substituting solutions (\ref{finalsol00})-(\ref{finalsol22}) into the field equations we arrive at the following relations for energy density and radial and tangential pressures: 
\bea
\rho(r)&=&\f{3{\sf Q}_0^2+8\pi r_0^2(\Lambda r_0^2-2)}{4\pi r_0^4(3{\sf w}_1-2{\sf w}_2-1)}\left(\f{r}{r_0}\right)^{\f{2({\sf w}_2-{\sf w}_1)}{{\sf w}_1}},\\\label{energy1}
p_{r}(r)&=&\f{{\sf w}_1\left[3{\sf Q}_0^2+8\pi r_0^2(\Lambda r_0^2-2)\right]}{4\pi r_0^4(3{\sf w}_1-2{\sf w}_2-1)}\left(\f{r}{r_0}\right)^{\f{2({\sf w}_2-{\sf w}_1)}{{\sf w}_1}},\\\label{pressurer}
p_{t}(r)&=&\f{{\sf w}_2\left[3{\sf Q}_0^2+8\pi r_0^2(\Lambda r_0^2-2)\right]}{4\pi r_0^4(3{\sf w}_1-2{\sf w}_2-1)}\left(\f{r}{r_0}\right)^{\f{2({\sf w}_2-{\sf w}_1)}{{\sf w}_1}}.\label{pressuret}
\eea
In {\sf GR}, the violation of {\sf NEC} is the basic requirement for the existence of wormhole solutions. The {\sf NEC} arises when one refers back to the Raychaudhuri equation, which is a purely geometric statement. Using the condition of the attractive nature of gravity for any hypersurface of orthogonal congruences (i.e., zero rotation
associated with the congruence defined by a null vector field) in these equations gives, ${\sf R}_{\mu\nu}k^\mu k^\nu\geq0$ and ${\sf R}_{\mu\nu}l^\mu l^\nu\geq0$ for null, $k^\nu$, and timelike, $l^\nu$, vector fields. If we replace the Ricci tensor with {\sf EMT} we obtain the standard energy conditions. Physical reliability requires that the wormhole configuration respects the {\sf WEC} and {\sf NEC} given by the following inequalities, respectively
\bea
\rho(r)\geq0,~~~~~\rho(r)+p_{r}(r)\geq0,~~~~~\rho(r)+p_{t}(r)\geq0,\label{wec}
\eea
\bea
\rho(r)+p_{r}(r)\geq0,~~~~~\rho(r)+p_{t}(r)\geq0.\label{nec}
\eea
We, therefor, require that the following conditions hold
\begin{enumerate}
\item The shape function satisfies the flare-out condition $rb^\prime-b<0$ which turns into inequality (\ref{ineqflare}) at the throat.
\item In order that the wormhole configuration be traversable, the spacetime must be free of horizons (the surfaces with ${\rm e }^{2\Phi(r)}\rightarrow0$); therefore the redshift function must be finite everywhere.
\item For asymptotic flat solutions, i.e., $\f{b(r)}{r}\rightarrow0$ as $r\rightarrow\infty$, we must have ${\sf w}_2/{\sf w}_1<0$. This condition also satisfies the inequality ${\sf w}_2/{\sf w}_1<1$, which is required for energy density and pressure profiles to converge asymptotically.
\item The square of the spin density must be positive at the throat and throughout the spacetime.
\item The coefficient of energy density must be positive so that the first inequality in (\ref{wec}) is satisfied.
\item ${\sf w}_1>-1$ and ${\sf w}_2>-1$ along with condition (3) so that the second and third inequalities in (\ref{wec}) are fulfilled. These conditions also guarantee the satisfaction of {\sf WEC} at the wormhole throat, i.e., $\rho(r_0)+p_r(r_0)>0$ and $\rho(r_0)+p_t(r_0)>0$. We also note that {\sf WEC} implies the null form.
\end{enumerate}
Figure (\ref{fig1}) shows the allowed values of ${\sf w}_1$  and ${\sf w}_2$ for which the above conditions are respected.
In figure (\ref{fig2}) we have plotted for the inverse of radial component of the metric (${\sf g}_{{\rm rr}}^{-1}(r)$) and the ratio $b(r)/r$. It is seen that the radial metric component stays positive for $r>r_0$ and thus the metric signature does not change. 
In figure (\ref{fig4}) we have plotted the electric field for fixed values of {\sf EoS} parameters but different values of charge at the throat. It is seen that the electric field gets different maximums for different values of ${\sf Q}_0$.  As the electric field reaches a maximum value around the throat, the charged particles will experience maximum attractive or repulsive forces depending on their sign. In figure (\ref{fig6}) we plotted for energy density, $\rho(r)+p_r(r)$ and $\rho(r)+p_t(r)$ where it is seen that the {\sf WEC} is satisfied for the obtained wormhole solutions.

\begin{figure}
\centering
\hbox{\hspace{1.11cm}\includegraphics[scale=0.45]{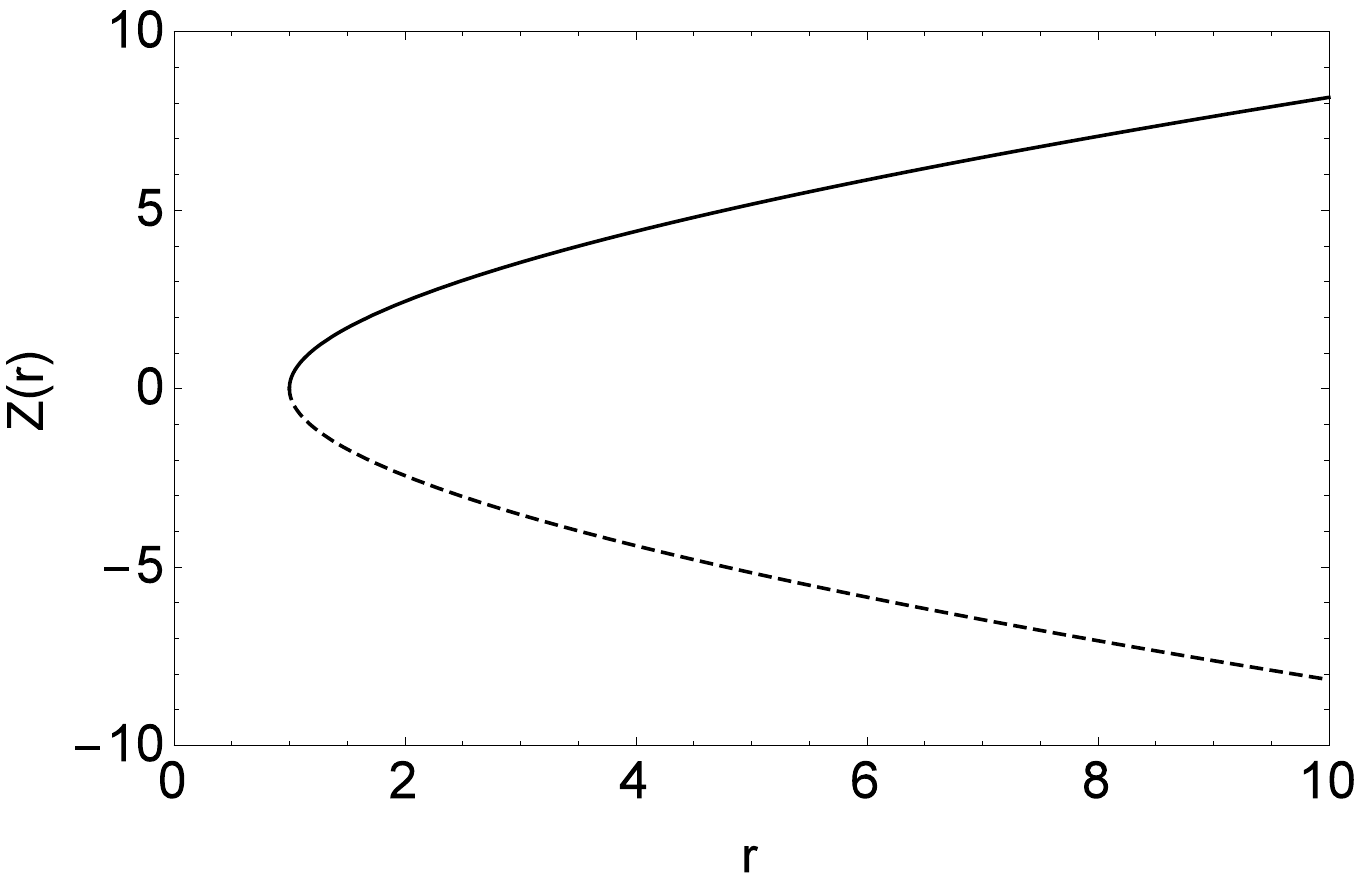}
\includegraphics[scale=0.50]{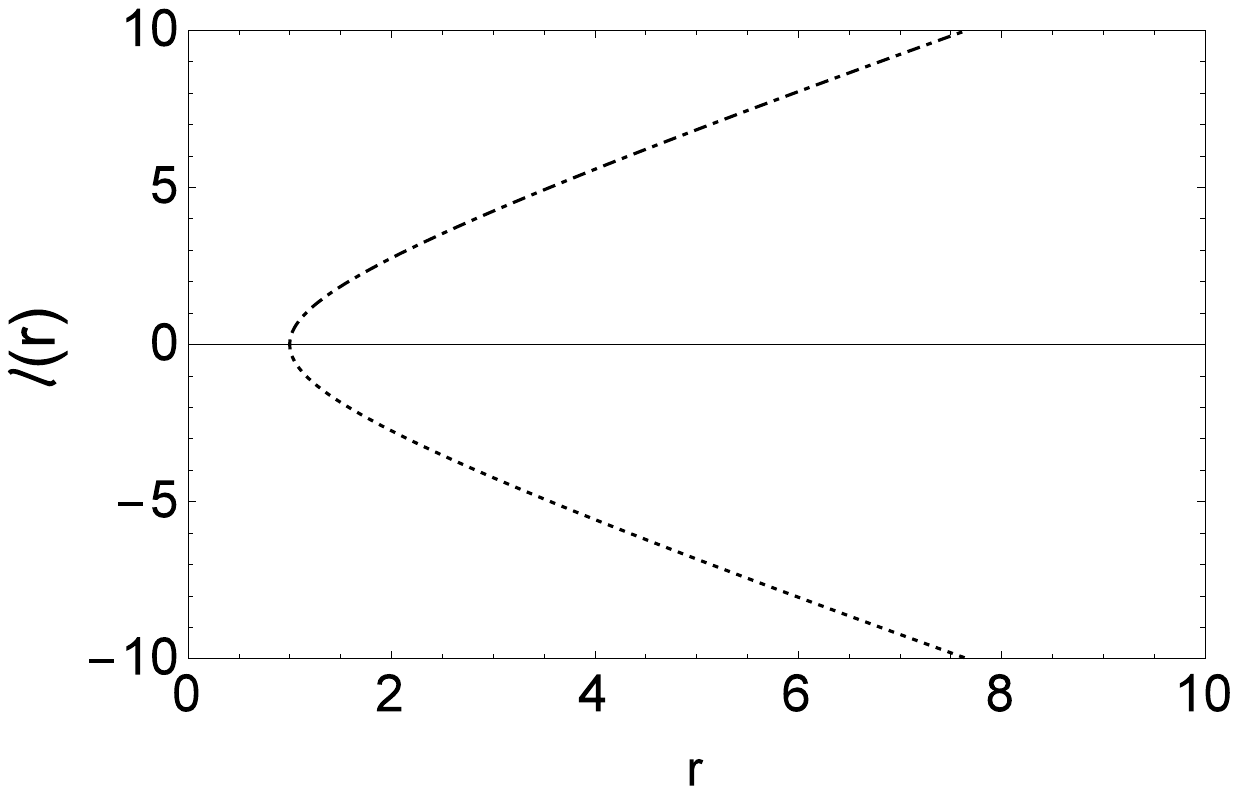}}
\caption{(Left panel) The embedding diagram of the wormhole (\ref{2dlinee}) obtained from solution (\ref{finalsol22}) for ${\sf w}_1 =-0.33$, ${\sf w}_2 =1$, $\Lambda=0$, $r_0=1$ and ${\sf Q}_0=1.1$. The solid and dashed lines correspond to $+$ and $-$ signs in (\ref{diffeqemdeb}), respectively. (Right panel) Plot of the radial proper distance of the wormhole against radial coordinate for the same parameters of the left panel.}\label{figzl}
\end{figure}

In order to visualize the wormhole, we consider a 2D slice of the metric (\ref{evw}) by setting $t={\sf constant}$ and $\theta=\pi/2$\footnote{Due to the spherically symmetry we can consider an equatorial slice, $\theta=\pi/2$, without loss of generality.}. The line element is then found as
\be\label{2dlinee}
ds^2=\f{dr^2}{1-\f{b(r)}{r}}+r^2d\phi^2.
\ee
According to Morris and Thorne~\cite{mt}, for visualizing the above slice, we embed the metric (\ref{2dlinee}) into a 3D Euclidean space for which the line element in cylindrical coordinates $(Z,r,\phi)$ is written as
\be\label{EUCLID}
ds^2=dZ^2+dr^2+r^2d\phi^2.
\ee
We note that in the 3D Euclidean space the embedded surface has equation $Z = Z(r)$, hence the line element of the surface can be written as
\be\label{lineeuclid}
ds^2=\left[1+\left(\f{dZ}{dr}\right)^2\right]dr^2+r^2d\phi^2,
\ee
whereby matching with the line element (\ref{2dlinee}) we get the following differential equation, and the corresponding integral, for the embedding surface of the wormhole
\be\label{diffeqemdeb}
\f{dZ}{dr}=\pm\left[\f{r}{b(r)}-1\right]^{-\f{1}{2}},~~~~~~~~~~~~ Z(r)=\pm\int_{r_0}^{r}\f{dy}{\sqrt{\f{y}{b(y)}-1}}.
\ee
From the above equation we observe that the geometry has the minimum radius $r_0=b(r_0)$ (wormhole\rq{}s throat) at which the embedded surface  is vertical, i.e., $dZ/dr{|_{r\rightarrow r_0}}\rightarrow\infty$. Moreover, the radial coordinate $r$ is illbehaved near the throat, however, the proper radial distance defined as
\be\label{properr}
\ell(r)=\pm\int_{r_0}^{r}\f{dy}{\sqrt{1-\f{b(y)}{y}}},
\ee
must be well behaved everywhere, i.e., $\ell(r)$ must be finite at all finite $r$, throughout the spacetime. The $\pm$ signs refer to the two asymptotically flat regions (as $\ell\rightarrow\pm\infty$ or equivalently  $r\rightarrow\infty$ then $b/r\rightarrow0$) which are connected by the wormhole throat. Next, we proceed to evaluate the above integrals using solution (\ref{finalsol22}). Unfortunately, the integration cannot be carried out analytically but we can perform it using numerical methods. The left panel in figure (\ref{figzl}) shows  the behavior of embedding function versus the radial coordinate for the same parameters of Fig. (\ref{fig2}). The embedding diagram shows that the wormhole extends from the throat located at $r=1$ to infinity. In the right panel we have plotted for proper radial distance of the wormhole. It is observed that as $\ell$ increases from $-\infty$ to zero the radial coordinate decreases monotonically to a minimum value at the wormhole\rq{}s throat; and as $\ell$ tends to $+\infty$, $r$ increases monotonically. 

\section{Observational Features}\label{OBSFEATURE}
\begin{figure}  
	\begin{center}
		\includegraphics[scale=0.2]{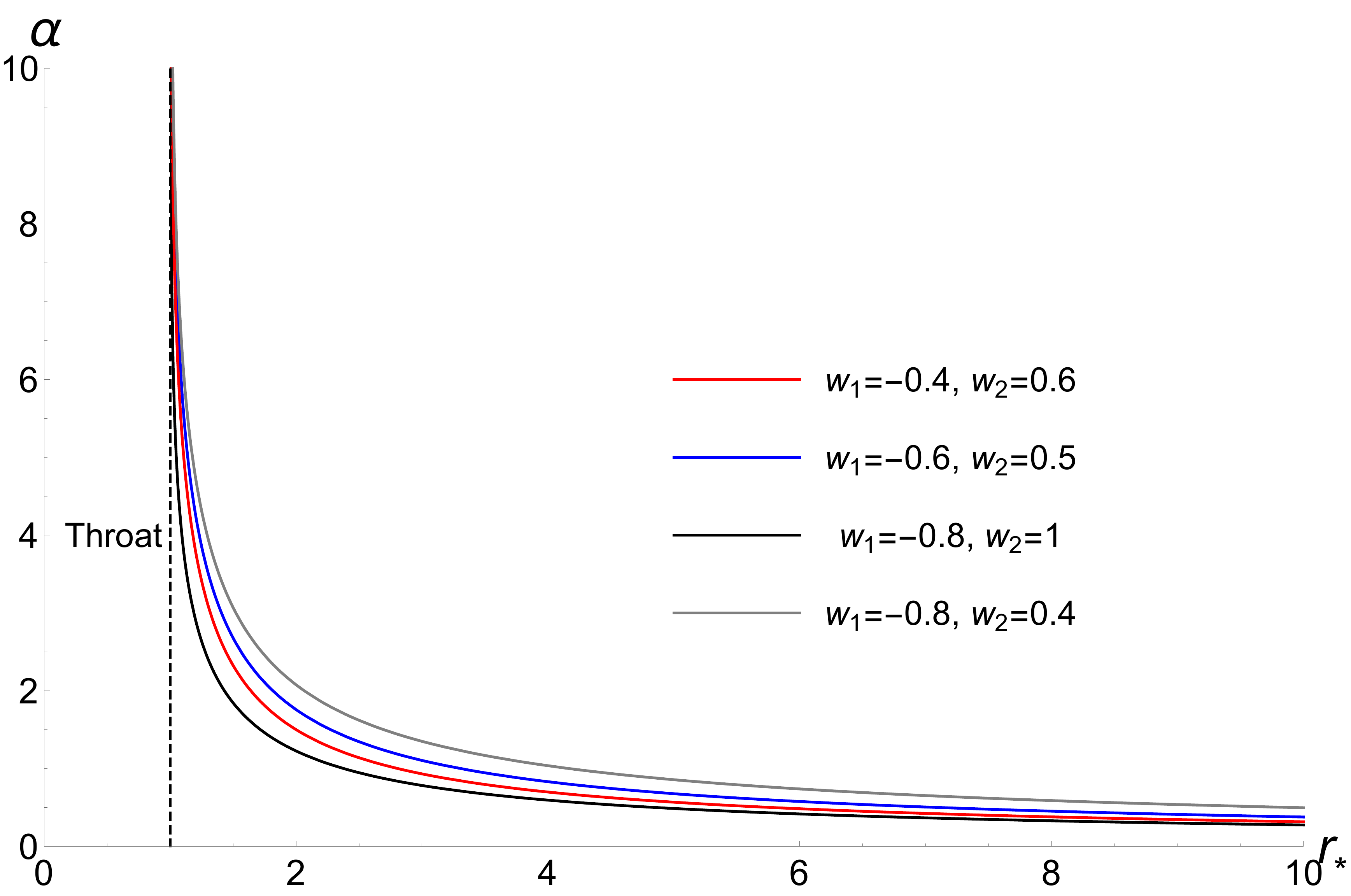}
		\caption{The deflection angle as a function of closest distance approach for ${\sf Q}_0=1.1$, $\Lambda=0$ and $r_0=1$. The {\sf EoS} parameters are picked up from the gray region of Fig. (\ref{fig1}).}\label{figdf}
	\end{center}
\end{figure}

A possible method for probing wormholes is the gravitational lensing effects. In the following we discuss lensing features of the wormhole spacetime presented in previous section. To that end we have to consider the lightlike geodesics. Because of spherical symmetry, it suffices to consider the equatorial plane $\theta=\pi/2$. The corresponding Lagrangian for the spacetime metric (\ref{evw}) then reads
\be\label{Lag}
{\mathcal L}=\f{1}{2}g_{\mu\nu}\dot{x}^\mu\dot{x}^\nu=\f{1}{2}\left[-{\rm e}^{2\Phi(r)}\dot{t}^2+\left(1-\f{b(r)}{r}\right)^{-1}\dot{r}^2+r^2\dot{\phi}^2\right],
\ee
where an overdot denotes derivative with respect to the curve parameter $\eta$. 
The Lagrangian ${\mathcal L}(\dot{x},x)$ is constant along a geodesic, so we can speak of timelike, lightlike and spacelike geodesics. Timelike geodesics (${\mathcal L}<0$) are to be interpreted as the world lines of freely falling particles, and lightlike ones (${\mathcal L}=0$) are to be interpreted as light rays. The geodesic equations can be obtained using the Euler-Lagrange equation
\be\label{EUEQ}
\f{d}{d\eta}\f{\partial{\mathcal L}}{\partial\dot{x}^\mu}-\f{\partial{\mathcal L}}{\partial x^\mu}=0,
\ee
which for Lagrangian (\ref{Lag}) reduces to the following differential equations 
\bea
&&\dot{t}={\mathcal E}{\rm e}^{-2\Phi(r)},\label{geoeqsLag}\\
&&r^2\dot{\phi}={h},\label{geoeqsLag0}\\
&&\left(1-\f{b(r)}{r}\right)^{-1}\left[\ddot{r}+\f{rb^\prime-b}{2r(r-b)}\dot{r}^2\right]+\Phi^\prime{\mathcal E}^2{\rm e}^{-2\Phi}-\f{h^2}{r^3}=0,\label{geoeqsLag1}
\eea
where ${\mathcal E}$ is the total energy of the particle in its orbit and $h$ is the particle's specific angular momentum. These equations are valid for both null and non-null geodesics. In each of these cases, however, it is easier to replace the rather complicated equation Eq. (\ref{geoeqsLag1}) by a first integral of geodesics equations which is nothing but the Lagrangian (\ref{Lag}). Using then Eqs. (\ref{Lag}), (\ref{geoeqsLag}) and (\ref{geoeqsLag0}) we can write the equation of photon trajectory as 
\be\label{effequation}
\dot{r}^2+{\rm e}^{-2\Phi(r)}\left(1-\f{b(r)}{r}\right)\left(V_{\rm eff}-{\mathcal E}^2\right)=0,~~~~V_{\rm eff}={\rm e}^{2\Phi(r)}\f{h^2}{r^2},
\ee
where $V_{\rm eff}$ is the effective potential. A light ray incoming from infinity will reach the closest approach distance $r_\star$ from the center of the gravitating
source and then emerge in another direction. Eliminating $dt$ from (\ref{effequation}) and (\ref{geoeqsLag0}) we get the deflection angle as a function of the closet approach~\cite{SWeinbergbook}
\be\label{defangleequa}
\alpha(r_\star)=-\pi+2\int_{r_\star}^{\infty}\f{{\rm e}^{\Phi(r)}dr}{r^2\left[\left(1-\f{b(r)}{r}\right)\left(\f{1}{\mu^2}-\f{{\rm e}^{2\Phi(r)}}{r^2}\right)\right]^{\f{1}{2}}},
\ee
where $\mu=h/{\mathcal E}$ is the impact parameter and at $r_\star$, $dr/d\phi$ vanishes so we have $\mu=r_\star{\rm e}^{-\Phi(r_\star)}$. For stationary observers, the tidal forces can be made to vanish by simply choosing $\Phi(r)=0$. Thus, for these ultra-static wormhole solutions we have
\be\label{defangleb}
\alpha(r_\star)=-\pi+2\int_{r_\star}^{\infty}\f{r_\star dr}{r\left[1-\f{3{\sf A}}{4\pi r_0^2}\left(\f{r_0}{r}\right)^{\f{2}{3}}+\f{{\sf w}_1({\sf w}_2-1){\sf B}}{4\pi r_0^2}\left(\f{r}{r_0}\right)^{\f{2{\sf w}_2}{{\sf w}_1}}\right]^{\f{1}{2}}\sqrt{r^2-r_\star^2}}.
\ee
As the impact parameter $\mu$ or correspondingly the closet distance approach, $r_\star=\mu$, decrease, the deflection angle grows. Decreasing $\mu$ further brings the light ray infinitesimally closer to the photon orbit causing the ray to wind up a large number of times before emerging out. Consequently, as $r_\star\rightarrow r_{\star}^{\rm m}$,  the deflection angle will diverge and the light ray will wind around a circular photon orbit indefinitely. In such a situation, circular photon orbits, that the collection of which constructs the photon sphere, satisfy $\dot{r}=\ddot{r}=0$. Moreover, stable (unstable) photon orbits meet the condition $\dddot{r}<0~(>0)$, respectively\footnote{In other words, a photon sphere is called stable if all lightlike geodesics with initial values close to those of a circular geodesic stay close to the photon sphere; otherwise it is called unstable~\cite{Perlicklvr}.}. In terms of the effective potential, these conditions are written as~\cite{Hasse-Perlick}
\be\label{CONDPHSHP}
V_{{\rm eff}}(r_{{\rm ph}})={\mathcal E}^2,~~~~\f{dV_{{\rm eff}}}{dr}\bigg|_{r=r_{{\rm ph}}}\!\!\!\!\!\!\!\!=0,~~~~~\f{d^2V_{{\rm eff}}}{dr^2}\bigg|_{r=r_{{\rm ph}}}\!\!\!\!\!\!\!\!>0~~~~{\rm or}~~~\f{d^2V_{{\rm eff}}}{dr^2}\bigg|_{r=r_{{\rm ph}}}\!\!\!\!\!\!\!\!<0,
\ee
where the inequalities correspond to stable and unstable (with respect to perturbations of the initial conditions) photon spheres, respectively. The equalities provide necessary and sufficient conditions for a null geodesic with circular orbit to exist at radius $r_{\rm ph}$. Using the second equality, the condition on existence of at lease one photon sphere requires that $r_{\rm ph}\Phi^\prime(r_{\rm ph})=1$ which obviously does not hold for the present wormhole configuration. However, there could exist circular photon orbits (either stable or unstable) at the wormhole throat where the effective potential exhibits an extremum. In order to better understand the situation, let us switch over from the Schwarzschild gauge to the proper radial coordinates defined by equation (\ref{properr}) where $\ell(r_0)=0$. The equation of photon trajectory then becomes
\be\label{eqphtrajell} 
\dot{\ell}^2+V_{{\rm eff}}-{\mathcal E}^2=0,~~~~~~~V_{{\rm eff}}=\f{h^2}{r^2(\ell)}.
\ee
Therefore, the first and second derivatives of effective potential at the throat are found as
\bea
\f{dV_{{\rm eff}}}{d\ell}\bigg|_{\ell=0}\!\!\!\!\!\!\!\!&=&\!\!\!\!\pm\left(1-\f{b(r)}{r}\right)^{\f{1}{2}}\f{dV_{{\rm eff}}}{dr}\bigg|_{r=r_0}\!\!\!\!\!\!\!\!=0,\label{fsdereffpot}\nn
\f{d^2V_{{\rm eff}}}{d\ell^2}\bigg|_{\ell=0}\!\!\!\!\!\!\!\!&=&\!\!\!\!\left(1-\f{b(r)}{r}\right)\f{d^2V_{{\rm eff}}}{dr^2}\bigg|_{r=r_0}+\f{1}{2}\left[\f{b(r)}{r^2}-\f{b^\prime(r)}{r}\right]\f{dV_{{\rm eff}}}{dr}\bigg|_{r=r_0}\nn
&=&\f{1-b^\prime(r_0)}{2r_0}\f{dV_{{\rm eff}}}{dr}\bigg|_{r_0},
\eea
where use has been made of the flare-out condition $b^\prime(r_0)<1$ along with $b(r_0)=r_0$. Using the second part of (\ref{effequation}) we finally get
\be\label{condsphsphere}
V_{{\sf eff}}(r_0)={\mathcal E}^2,~~~~~\f{dV_{{\rm eff}}}{d\ell}\bigg|_{\ell=0}\!\!\!\!\!\!\!\!=0,~~~~~~\f{d^2V_{{\rm eff}}}{d\ell^2}\bigg|_{\ell=0}\!\!\!\!\!\!\!\!<0,
\ee
whence we observe that in proper radial coordinate, the effective potential admits a maximum at the throat so the wormhole throat acts as an effective photon sphere located at $r_{\star}^{\rm m}=r_0$. This can also be seen by evaluating the deflection angle given by integral (\ref{defangleb}). However, this integral cannot be expressed in closed form in terms of known special functions and one may resort to an approximate analytical solution. Introducing new variable $x=1-r_\star/r$, the integral can be rewritten as
\be\label{newvarxint}
\!\!\!\!\!\!\!\!\!\alpha(\epsilon)=-\pi+2\int_{0}^{1}\f{\left[\left(1-\f{3{\sf A}}{4\pi r_0^2}\epsilon^{\f{2}{3}}(1-x)^{\f{2}{3}}+\f{{\sf w}_1({\sf w}_2-1){\sf B}}{4\pi r_0^2}\epsilon^{-\f{2{\sf w}_2}{{\sf w}_1}}(1-x)^{-\f{2{\sf w}_2}{{\sf w}_1}}\right)\right]^{-\f{1}{2}}}{\sqrt{(2x-x^2)}}dx,~~~~~\epsilon=\f{r_0}{r_\star}.
\ee
For a special case where ${\sf A}=0$, the square of charge function at the throat is found as
\be\label{sqq0}
{\sf Q}_0^2=\f{4\pi r_0^2({\sf w}_2-{\sf w}_1)(1+{\sf w}_1+2{\sf w}_2)}{{\sf w}_1({\sf w}_2-1)},
\ee
from which we get ${\sf B}=4\pi r_0^2/({\sf w}_1(1-{\sf w}_2))$. For ${\sf Q}_0\in\mathbb{R}$ together with considering the gray region in Fig.(\ref{fig1}) we require that $-1<{\sf w_1}<0$ and $0<{\sf w_2}<1$. We note that care should be taken for choosing appropriate values for {\sf EoS} parameters in order to avoid a vanishing denominator in expression (\ref{coeff0}). Therefore, in the weak field limit where $\epsilon\ll1$ we can expand the numerator in the integrand up to second order to get
\be\label{expandnumin}
\alpha(\epsilon)\approx-\pi+2\epsilon^{-\f{2{\sf w}_2}{{\sf w}_1}}\int_{0}^{1}\f{1+\f{1}{2}(1-x)^{-\f{2{\sf w}_2}{{\sf w}_1}}}{\sqrt{2x-x^2}}dx+{\mathcal O}(\zeta^2),~~~\zeta=\epsilon^{-\f{2{\sf w}_2}{{\sf w}_1}}.
\ee
The integration can be performed giving
\be\label{alphaepsilonapp}
\alpha(r_\star)\approx\sqrt{\pi}\left(\f{r_0}{r_\star}\right)^{-\f{2{\sf w}_2}{{\sf w}_1}}\f{\Gamma\left[1-\f{2{\sf w}_2}{{\sf w}_1}\right]}{\Gamma\left[\f{3}{2}-\f{2{\sf w}_2}{{\sf w}_1}\right]}{}_2 {\sf F}_1\left[\f{1}{2},1-\f{2{\sf w}_2}{{\sf w}_1},\f{3}{2}-\f{2{\sf w}_2}{{\sf w}_1},-1\right].
\ee
From the above approximation, we observe that the deflection angle in the limit where ${\sf w}_1\rightarrow-{\sf w}_2$ tends to that of Ellis wormhole in weak field regime~\cite{TsukamotoHaradaYajima2012}. As the closet distance approach increases, the deflection angle decreases, however, we are not allowed to evaluate the above expression in the limit of approach to the throat as these formula is valid in the weak field regime where $r_0\ll r_\star$. Nevertheless, the expression (\ref{defangleb}) can be fully integrated using numerical methods. By so doing, we obtain numerical solution for deflection angle in terms of closest distance approach, as shown in Fig.~(\ref{figdf}). When the turning point $r_\star$ is away from the photon sphere, $\alpha(r_\star)$ is a nonzero positive finite number and thus, the deflection suffered by the light ray is finite. The deflection tends to zero as $r_\star\rightarrow \infty$, that is, no scattering for the light ray occurs. Furthermore, the deflection angle grows as $r_\star\rightarrow r_0$ and diverges at the wormhole throat where an unstable photon sphere is present. Consequently, the wormhole configuration can produce infinitely many relativistic images of an appropriately placed light source. This infinite sequence corresponds to infinitely many light rays whose limit curve asymptotically spirals towards the unstable photon sphere~\cite{Hasse-Perlick}. Since the photon sphere is located at the wormhole throat, such a sphere may be detectable (by highly sensitive instruments) thus providing observational evidence for the existence of the wormhole. Beside the lensing effects, different observational features have been studied so far with the aim of probing a wormhole configuration which inhabits our Universe. Work along this line includes the investigation of particle trajectory in the vicinity of the wormhole structure~\cite{parttrajworm}, accretion disks around wormholes~\cite{accrdiskworm} and their gravitational wave signatures~\cite{Gwaveworm}. Another observational aspect which can be utilized to
extract physical information is the shadows cast by a wormhole or its apparent shape~\cite{wormshadow}. The appearance of a shadow is a phenomenon which does not belong only to black hole spacetimes and under some conditions this interesting event can be observed also by other compact objects such as wormholes. Motivated by this idea, Ohgami and Sakai~\cite{OhgamiSakai2005} have recently developed the shadowlike images of wormholes surrounded by optically thin dust and a shadow cast by rotating wormholes has been studied in~\cite{raotwormshadow}. The presence of unstable photon orbits could play a crucial role in studying wormhole shadows as these orbits define the boundary between capture and noncapture of the light rays around a wormhole. Thus, the boundary of the shadow is only determined by the metric of spacetime since it corresponds to the apparent shape of the photon sphere as seen by a distant observer~\cite{raotwormshadow,Falcke2000} (see also~\cite{GRGSHADOWGL} for a recent review on shadows). Hence, gravitational lensing effects along with shadows are of remarkable importance as they will allow us to extract useful information from the wormhole structure and the interactions of the wormhole with its astrophysical environment.

\section{Concluding Remarks}\label{concluding}
In the present work we constructed models of static wormholes within the framework of  {\sf ECT} by considering usual (nonexotic) spinning matter (Weyssenhoff fluid) along with an anisotropic energy momentum tensor ({\sf EMT}) and a Maxwell field as supporting matters for the wormhole geometry. The radial and tangential components for anisotropic fluid pressures were taken to depend linearly to energy density via different {\sf EoS} parameters. These {\sf EoS} parameters along with the value of the charge and spin square density at the throat construct a space of parameters. The allowed regions of this parameter space are subject to fulfillment of physical conditions on wormhole configuration, a subspace of which is sketched in Fig. (\ref{fig1}). In general, our solutions include the ranges ${\sf w}_1<0$ and ${\sf w}_2>0$ for {\sf EoS} parameters, that is the wormhole configurations are supported by positive and negative pressures along lateral and radial directions, respectively. Thus the anisotropy parameter, which for our model is given by $\Delta(r)=p_t-p_r=({\sf w}_2-{\sf w}_1)\rho(r)$ is always positive unless   the {\sf WEC} is violated. This implies that the geometry is repulsive due to the anisotropy of the system. From the viewpoint of the equilibrium condition for wormhole configuration we can divide conservation equation (\ref{conseq}), which is the well-known Tolman-Oppenheimer-Volkov equation, into three parts~\cite{PONPoisson}: the anisotropic force $F_{\sf a}=2\Delta/r$, the hydrostatic force $F_{\sf h}=-p_r^\prime$ and the force due to gravitational contribution $f_{\sf g}=-\Phi^\prime[\rho(r)+p_r(r)]$, where the last one is absent in our model since the redshift function is constant.  Taking the derivative of radial pressure, it is easy to check that $F_{\sf h}=-F_{\sf a}$ and hence we can deduce that the wormhole solutions are in equilibrium as the anisotropic and hydrostatic forces cancel each other~\cite{anisorepul}. As Fig. (\ref{fig4}) shows, the electric field admits a maximum near the throat and this provides a setting to accelerate charged particles toward wormhole, namely, the more a negative charge approaches the wormhole throat the more attractive force (${\sf E}(r)>0$) it feels; this scenario also occurs for positive charges considering a negative sign for charge function. Beside the present model,  wormhole solutions in the presence of electric charge have been studied earlier. For example, by adding an electric charge, the authors studied the possibility of stabilizing a wormhole supported by a ghost scalar field~\cite{Gonzales- Guzman}. Charged wormholes in Einstein-Maxwell theory have been constructed by real feasible matter sources in~\cite{F. Rahaman2007} where the solutions respect the energy conditions throughout the spacetime but the {\sf NEC} is violated at the throat. Finally, we would like to remark that, though the effects of spacetime torsion has not been observed yet, one may not definitely decide for irrelevance of torsion interactions in gravitational physics, since, for example, the interaction of matter fields with torsion may not be negligible in the realm of particle physics~\cite{shapirophysrep}. In cosmological models based on {\sf ECT} (see e.g.~\cite{Gas}), it is shown that the spin degrees of freedom of fermionic matter plays a significant role in cosmological evolution of the very early Universe where the square of spin density is dominant and is scaled as $(1+z)^6$~\cite{spinfluidz}, with $z$ being the cosmological redshift. As the Universe expands, the spin contribution becomes diluted and is no longer dominant over the radiation term before the big bang nucleosynthesis (BBN) commences. However, by imposing BBN and cosmic microwave background constraints, a limit of present value for the density parameter of spin fluid is found as $\Omega_{{\sf 0s}}=-0.012$ at the $1\sigma$ level~\cite{spinfluidz,spinfluidz1}.
\par 
In conclusion, if we assume that a gas of charged fermionic particles could be capable of producing a wormhole configuration, then, from the dependency of the throat radius or correspondingly the radius of photon sphere on $({\sf S}_0^2,{\sf Q}_0^2)$, we may deduce that the pattern of lensing images and possibly the apparent shape of the wormhole can be affected by spin and charge distribution at the throat providing thus a setting to search for possible footprints of spacetime torsion in nature.
\section{Acknowledgements} 
The authors would like to thank the anonymous referee for comments and suggestions that helped improve the manuscript. The work of A. H. Ziaie has been supported
financially by Research Institute for Astronomy \& Astrophysics of Maragha (RIAAM).

\end{document}